%% file: p.tex
\documentclass[sigconf,dvipsnames,screen]{acmart}

\usepackage{booktabs} % For formal tables
\usepackage{xcolor}
\usepackage{soul}

\usepackage{siunitx}
\usepackage{multirow}
\usepackage{balance}
\usepackage[font=rm]{caption}
\usepackage{algorithm}
\usepackage[noend]{algpseudocode}
\usepackage{verbatim}
\usepackage{graphicx}
\usepackage{tabularx}
\usepackage{enumitem}
\usepackage{shortvrb}
\usepackage{adjustbox}
\setlist[itemize,1]{leftmargin=3mm,itemsep=1mm}
\setlist[enumerate,1]{leftmargin=3mm,itemsep=1mm}

\input{macros.tex}
\setcopyright{acmlicensed}
\copyrightyear{2025}
\acmYear{2025}
\setcopyright{rightsretained}
\acmConference[SIGIR '25] {Proceedings of the 48th International ACM
SIGIR Conference on Research and Development in Information
Retrieval}{July 13--18, 2025}{Padua, Italy.}
\acmBooktitle{Proceedings of the 48th International ACM SIGIR
Conference on Research and Development in Information Retrieval
(SIGIR '25), July 13--18, 2025, Padua, Italy}
\acmISBN{979-8-4007-1592-1/25/07}
\acmDOI{10.1145/3726302.3730255}

\makeatletter \gdef\@copyrightpermission{
\begin{minipage}{0.3\columnwidth}%
\href{https://creativecommons.org/licenses/by/4.0/}%
{\includegraphics[width=0.90\textwidth]{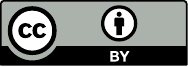}}%
\end{minipage}\hfill%
\begin{minipage}{0.7\columnwidth}
\href{https://creativecommons.org/licenses/by/4.0/}%
{This work is licensed under a Creative Commons Attribution
International 4.0 License.}%
\end{minipage}%
\vspace{5pt}}%
\makeatother

\begin{document}
%% \fancyhead{}
\title{The Effects of Demographic Instructions on LLM Personas}

%% Authorship order discussed at meeting on 30 Jan with all of
%% Falk, Shane, Johanne, Angel, Sachin, and Alistair present an
%% in agreement that it will be: Angel first and then alpha by
%% family name, that is:
%% Angel Felipe Magnossão de Paula
%% J . Shane Culpepper
%% Alistair Moffat
%% Sachin Pathiyan Cherumanal
%% Falk Scholer
%% Johanne R. Trippas
%% and also agreed thet there are no other people eligible to be
%% considered as authors

\author{Angel Felipe Magnossão de Paula}
\orcid{0000-0001-8575-5012}
\affiliation{%
  \institution{Universitat Politècnica de València}
  \city{València}
  \country{Spain}
}
\email{adepau@doctor.upv.es}

\author{J. Shane Culpepper}
\orcid{0000-0002-1902-9087}
\affiliation{%
  \institution{University of Queensland}
  \city{Brisbane}
  \country{Australia}
}
\email{s.culpepper@uq.edu.au}

\author{Alistair Moffat}
\orcid{0000-0002-6638-0232}
\affiliation{%
  \institution{The University of Melbourne}
  \city{Melbourne} 
  \country{Australia} 
}
\email{ammoffat@unimelb.edu.au}

\author{Sachin Pathiyan Cherumanal}
\orcid{0000-0001-9982-3944}
\affiliation{%
  \institution{RMIT University}
  \city{Melbourne} 
  \country{Australia} 
}
\email{s3874326@student.rmit.edu.au}

\author{Falk Scholer}
\orcid{0000-0001-9094-0810}
\affiliation{%
  \institution{RMIT University}
  \city{Melbourne} 
  \country{Australia} 
}
\email{falk.scholer@rmit.edu.au}

\author{Johanne Trippas}
\orcid{0000-0002-7801-0239}
\affiliation{%
\institution{RMIT University}
\city{Melbourne}
\country{Australia}
}
\email{j.trippas@rmit.edu.au}

\renewcommand{\shortauthors}{Angel Felipe Magnossão de Paula et al.}

\input{sec0-abstract.tex}


%% \ccsdesc[300]{Human-centered computing~User models}
%% \ccsdesc[500]{Information systems~Presentation of retrieval results}
%% \ccsdesc[100]{Information systems~Search engine architectures and scalability}


\maketitle

\input{sec1-intro.tex}
\input{sec2-background.tex}

\input{sec3-methodology}

\input{sec4-results}

\input{sec5-conclusion}

%% \input{LINKS}

%-------------------------------------------------------------------%

\myparagraph{Acknowledgment}
This work was supported in part by the {\grantsponsor{ARC}{Australian
Research Council}{https://www.arc.gov.au/}} (projects
{\grantnum{ARC}{DP190101113}}, {\grantnum{ARC}{DE200100064}}, and
{\grantnum{ARC}{CE200100005}}) and was undertaken with the assistance
of computing resources from RACE (RMIT AWS Cloud Supercomputing).
%% \alistair{Others??} \jt{none from me}

% \myparagraph{Software and Data}
% \alistair{Submission said that these would be provided.
% Now need a URL.
% Two lines max!}

%% \myparagraph{Software}
\renewcommand{\bibsep}{3.5pt}
\bibliographystyle{abbrvnat}
\balance %% <- doesn't work with footnotes
\bibliography{strings-shrt,local-trimmed} 

\end{document}

%% file: macros.tex
\newcommand{\method}[1]{{\sf{#1}}}

\newcommand{\myparagraph}[1]{\paragraph*{\hspace*{-\parindent}\normalsize\bf#1}}
\newcommand{\mycaption}[1]{\caption{{\rm{#1}}}}

\usepackage{xcolor}
\definecolor{tealgreen}{rgb}{0.0, 0.51, 0.5}

\newcommand{\strike}[1]{}

\newcommand{\ir}{\text{IR}}
\newcommand{\llm}{\text{LLM}}
\newcommand{\llms}{\text{LLMs}}

\newcommand{\gptthree}{\method{GPT-3.5}}
\newcommand{\gptfour}{\method{GPT-4}}
\newcommand{\gptl}{\method{GPT-4o}}
\newcommand{\mistral}{\method{Mistral}}
\newcommand{\qwen}{\method{Qwen}}

\newcommand{\gptthreefemale} {$\text{\gptthree}_{F}$}
\newcommand{\gptthreemale} {$\text{\gptthree}_{M}$}

\newcommand{\gptfourfemale} {$\text{\gptfour}_{F}$}
\newcommand{\gptfourmale} {$\text{\gptfour}_{M}$}

\newcommand{\gptlfemale} {$\text{\gptl}_{F}$}
\newcommand{\gptlmale} {$\text{\gptl}_{M}$}

\newcommand{\mistralmale} {$\text{\mistral}_{M}$}
\newcommand{\mistralfemale} {$\text{\mistral}_{F}$}

\newcommand{\qwenmale} {$\text{\qwen}_{M}$}
\newcommand{\qwenfemale} {$\text{\qwen}_{F}$}

\newcommand{\gptthreeageone} {$\text{\gptthree}_{18-22}$}
\newcommand{\gptfourageone} {$\text{\gptfour}_{18-22}$}
\newcommand{\gptlageone} {$\text{\gptl}_{18-22}$}
\newcommand{\mistralageone} {$\text{\mistral}_{18-22}$}
\newcommand{\qwenageone} {$\text{\qwen}_{18-22}$}

\newcommand{\gptthreeagetwo} {$\text{\gptthree}_{23-45}$}
\newcommand{\gptfouragetwo} {$\text{\gptfour}_{23-45}$}
\newcommand{\gptlagetwo} {$\text{\gptl}_{23-45}$}
\newcommand{\mistralagetwo} {$\text{\mistral}_{23-45}$}
\newcommand{\qwenagetwo} {$\text{\qwen}_{23-45}$}

\newcommand{\gptthreeagethree} {$\text{\gptthree}_{46+}$}
\newcommand{\gptfouragethree} {$\text{\gptfour}_{46+}$}
\newcommand{\gptlagethree} {$\text{\gptl}_{46+}$}
\newcommand{\mistralagethree} {$\text{\mistral}_{46+}$}
\newcommand{\qwenagethree} {$\text{\qwen}_{46+}$}

\usepackage{listings}

%% file: sec0-abstract.tex
\begin{abstract}
Social media platforms must filter sexist content in compliance with
governmental regulations.
Current machine learning approaches can reliably detect sexism based
on standardized definitions, but often neglect the subjective nature
of sexist language and fail to consider individual users'
perspectives.
To address this gap, we adopt a perspectivist approach, retaining
diverse annotations rather than enforcing gold-standard labels or
their aggregations, allowing models to account for personal or
group-specific views of sexism.
Using demographic data from Twitter, we employ large language models
(\llms) to personalize the identification of sexism.

Our empirical results show that OpenAI's {\llms}
({\method{GPT-3.5}}, {\method{GPT-4}}, and {\method{GPT-4o}}) and two
open-source {\llms} ({\method{Mistral}} and {\method{Qwen}}) exhibit
higher Krippendorff's alpha label agreement with female annotators
than with male annotators.
As well, each {\llm} presents higher Krippendorff's alpha agreement
with a specific annotator age group.
We then sought to counter these trends by providing ``persona''
instructions as part of the {\llm} prompt, with somewhat surprising
outcomes, highlighting the potential of user-centered
perspectivist methods to improve content moderation systems.

\end{abstract}

%% file: sec1-intro.tex
%------------------------------------------
\section{Introduction and Motivation}
\label{sec-intro}
%------------------------------------------

Relevance judgments are usually undertaken to develop a set of gold-standard labels, seeking
relevance consensus; with those labels then used to support
measurement of the extent to which the documents retrieved by some
information retrieval ({\ir}) system align with users' information
needs, thereby allowing benchmarking of systems.
One major challenge in relevance judgments is the high annotator
costs for the labeling that is required.
To reduce that cost, researchers have proposed the use of {\llms},
which have been shown to usefully supplement human labeling and at
scale {\citep{thomas2024LLmindustry}}.

Bias and fairness in {\ir} systems have also received attention
{\citep{dai2024biasirllmeratutorial, lunardi2024elusiveness}}.
Nor are {\llms} immune -- they too can be affected by bias,
stereotypical associations
{\citep{basta-etal-2019-evaluating,kurita-etal-2019-measuring}}, and
adverse sentiments towards specific groups
{\cite{hutchinson-etal-2020-social}}.
For example, gender bias evaluation in natural language processing is
a topic that has received much attention
{\citep{costa2019proceedings}}, with de-biasing techniques having
been also been proposed {\citep{bigdeli2023debiasing}}.

{\citet{faggioli2023perspectivesllmjudgement}} propose a
human-machine collaboration spectrum that categorizes judgment
strategies based on how much humans rely on machines, and suggest
that ``AI Assistance'' is a likely path for employment of {\llms}.
{\citeauthor{faggioli2023perspectivesllmjudgement}}'s pilot study
finds a reasonable correlation between highly-trained human assessors
and a fully automated {\llm}, concluding that while the technology is
promising, it requires further study.
Use of {\llms} for relevance judgment is thus an emerging area of
interest {\citep{rahmani2024LLM4Eval, schnabel2025multistage}}.

Here, we explore if {\llms} can be directed to adopt a
{\emph{persona}} when conducting relevance judgments.
In particular, different annotators might react differently when
asked ``is this tweet sexist'', with their answers coming from
subjective viewpoints influenced by, amongst other factors, gender
and age.
That is, the ``is it sexist'' question may not always have a single
``right'' answer.
{\citet{hasler2015augmented}} suggest the use of augmented test
collections that include user-centric evaluation and anonymized
demographic information such as age, gender, and education level,
plus task-specific details such as the assessor's expertise,
interest, motivation, confidence, and degree of document relevance.
Given that range of influencing factors, we are interested in whether
an {\llm} -- a ``stochastic parrot''
{\citep{bender2021stochasticparrots}} -- is capable of reflecting
different demographic responses.
For example, can an {\llm} be (reliably) instructed to ``be a male
over~45''?

To investigate if {\llms} can mimic subjectivity, we use the Social
neTworks ({\method{EXIST}}) Shared Task at {\method{CLEF}} 2023
{\cite{lplaza2023existoverview}} test collection for sEXism
Identification.
This dataset contains labels (opinions) from diverse human judges for
a set of tweets\footnote{User-provided ground truth labels are not
publicly available, meaning that this dataset could not have been
used as training by {\llm}, an important and necessary assurance.},
labels which do indeed exhibit demographic patterns.
By comparing the human labels with {\llm}-generated labels, we are
thus able to infer a ``persona'' for several commercial
and open-source {\llm} models.
The specific research questions we consider are:
\smallskip

\noindent
({i})
Do {\llms} exhibit bias toward certain demographics when classifying
text as sexist or non-sexist; 
and
\smallskip

\noindent
({ii}) Can adopting a demographic-based persona mitigate bias
in {\llms} when classifying text as sexist or non-sexist.
\smallskip

\noindent
{\citet{krieg2022perceivedgenderbiasreljudgements}} suggest that
female stereotypes are influential in relevance judgments.
Our experiments show a similar outcome, with all five tested {\llms}
correlating more closely with female opinions than with male.
The second part of our work here then asks the same set of {\llms} to
adopt a range of specific demographic personas as they respond to the
``is this tweet sexist'' question.
Surprisingly, all of the {\llms} tested seemed to be incapable of
doing so.
That is, the {\llms} were unable to ``empathize'' and take on
different patterns of opinion; and their ``personalities'' seem to be
relatively inflexible.

%% file: sec2-background.tex
%------------------------------------------
\section{Background}
\label{sec-background}
%------------------------------------------

\smallskip

%------------------------------------------
\myparagraph{Perspectivism in Text Classification}
%------------------------------------------

In traditional text classification annotation, disagreements are
resolved into a single ``gold standard'' label via aggregation
{\cite{frenda2024perspectivist}}.
This approach has recently been challenged, especially in subjective
tasks like hate speech and sexism detection
{\cite{lplaza2023existoverview}}, as it risks enforcing a single
ground truth.
Perspectivism is a machine learning approach that takes data
annotated by different individuals and models the varied perspectives
that influence their opinions and world view
{\cite{frenda2024perspectivist}}.
We adopt the perspectivist approach to examine biases in sexism
detection; and then consider the specific question to whether
{\llms} can simulate personas based on demographic attributes.

%------------------------------------------
\myparagraph{Sexism Detection}
%------------------------------------------

Sexism detection is the task of deciding if text contains sexist
content.
Traditional sexism detection systems have relied on predefined labels
and fixed perspectives, overlooking the nuanced and subjective
nature of sexist statements; moreover, as social media have expanded
their influence, researchers have focused on developing scalable
approaches to sexism detection.

A significant advancement towards addressing this issue is the
{\method{EXIST}} (sEXism Identification in Social neTworks) Lab at
{\method{CLEF}} {\cite{lplaza2023existoverview}}, designed for
perspectivist learning by highlighting annotation disagreements
rather than imposing gold-standard labels.
In particular, different annotators might react differently when
asked if a given tweet ``is sexist''.
The {\method{EXIST}} initiative acknowledges the inherent
subjectivity in sexism classification, and aims to improve model
robustness by incorporating diverse perspectives in the annotation
process.
Various approaches have been proposed, ranging from rule-based
methods {\cite{samory2021call}} to machine learning techniques
{\cite{de2022detection,de2021sexism,9281090}}.
In this study we investigate the role of {\llms}.

%------------------------------------------
\myparagraph{LLM Personas}
%------------------------------------------

Recent research has explored how {\llms} can mimic personas based on
prompts that describe the demographics of a target user group using
{\emph{persona prompting}} {\cite{zhan2024unveiling}}.
However, the use of persona prompting is still poorly understood, and
has led to somewhat inconsistent outcomes, in differences that might
be attributed to the lack of clarity on how small edits to the input
prompt can produce unexpected changes in the generated text
{\cite{zhan2024unveiling,sclar2023quantifying}}.

To mitigate this issue, {\citet{aguda-etal-2024-large}} proposed a
``reliability index'' called {\method{{\llm}-Relindex}}, which can be
used to identify input prompts that may require a domain expert to
review the output results.
{\citeauthor{aguda-etal-2024-large}} found that
{\method{{\llm}-Reindex}} was most reliable when prompts were
customized by persona.
Furthermore, {\llm}-generated personas have also shown to exhibit
demographic biases.
{\citet{salminen2024duesexpersonas}} identified biases in age and
occupation, and a strong tendency towards personas from the United
States.
Furthermore, {\citet{zheng-etal-2024-helpful}} demonstrate that such
personas do not consistently improve performance; but that gender,
and domain-based personas do sometimes lead to improved performance.

%% file: sec3-methodology.tex
%------------------------------------------
\section{Methodology}
\label{sec-methodology}
%------------------------------------------

\smallskip

%------------------------------------------
\myparagraph{Data}
%------------------------------------------

We use the {\method{EXIST}} 2023 shared task
dataset~\cite{lplaza2023existoverview, plaza2023exist}, designed for
perspectivist learning by capturing annotation disagreements rather
than assigning single labels.
The collection focuses on sexism in tweets, and includes demographic
data about the annotators, enabling further analysis of bias and
subjectivity in the classification process.
The primary task involves distinguishing sexist from non-sexist
content, rated using a binary scale.

Each tweet is annotated by six individuals, stratified across two
factors: gender (male and female) and age group ($18$--$22$,
$23$--$45$, and $46$+), ensuring diversity of perspective.
There are $7{,}958$ tweets in total, divided into training,
development, and test sets.
We merged the training and development subsets into a single file to
increase the number of samples, and streamline analysis.

Instead of a rigid binary classification, a soft-labeling framework
based on the proportion of human annotators that selected each
category is used, capturing annotator disagreement by providing
probabilistic distributions over the two categories, which sum to
$1.0$.

The organizers of {\method{EXIST}} 2023 provided the annotations for
the training and development sets.
Table~\ref{tab:exist2023_data_distribution} presents the distribution
of annotations (sexist or non-sexist) across gender and age groups.
The dataset maintains a balanced composition of male and female
annotators, as well as across different age groups.

\input{Tables/tbl_exist2023_distribution}
%------------------------------------------

We analyze label distributions by demographic groups, and test for
statistically significant differences between group means using a
$t$-test for gender (two levels), and a one-way ANOVA for age-group
(three levels), against an alpha of $0.05$.
A post hoc Tukey's HSD test was performed to determine the specific
groups contributing to these differences.
Differences in the gender factor were not significant ($p=0.237$).
The ANOVA indicated a significant difference between the mean values
of different age groups, with the follow-up Tukey's HSD test showing
significant differences between the $18$--$22$ and the $23$--$45$
groups, and between the $18$--$22$ and the $46$+ groups, reinforcing
the presence of annotation differences: the $18$--$22$ age group
identified sexist content less frequently (and non-sexist content
more frequently) than the other groups, while the $23$--$45$ and
$46$+ groups exhibited more similar annotation patterns. These results highlight the role of the perspectivist approach.

%------------------------------------------
\myparagraph{Large Language Models}
%------------------------------------------

Three {\llms} from OpenAI were used to evaluate model performance in
detecting sexism: {\method{GPT-3.5}} ({\method{gpt-3.5-turbo-0125}}),
{\method{GPT-4}} ({\method{gpt-4-turbo-2024-04-09}}), and
{\method{GPT-4o}} ({\method{gpt-4o-2024-08-06}}).
Additionally, we included two open-source {\llms}, {\mistral}
({\method{Mistral-Small-Instruct-2409 22B}}) and {\qwen}
({\method{Qwen2.5-14B}}), to provide a comparative analysis of sexism
detection capabilities across these five model architectures.

%------------------------------------------
\myparagraph{Prompt Creation}
%------------------------------------------

Three prompt candidates were developed based on the {\method{EXIST}}
2023 task guidelines,\footnote{\url{https://nlp.uned.es/exist2023/}}
focusing on specificity, grammar, and clarity.
These three prompts were used to annotate twenty randomly selected
tweets from the {\method{EXIST}} 2023 dataset.
The prompt with the highest output consistency across three {\llms}
was chosen, where consistency was measured as the percentage of cases
in which all models produced the same classification.\footnote{With
small optimization changes required to employ the prompts against the
two open source models, {\method{Mistral}}, and {\method{Qwen}},
both of which rank highly on leaderboards for their model size.}
The prompt with the highest output consistency achieved a $75$\%
success rate, followed by the other two prompts, which attained
$70$\% and $55$\%.

Following the lead of others {\cite{DBLP:conf/iclr/ZhouMHPPCB23,
kepel2024autonomouspromptengineeringlarge}}, we optimized the prompt
using an {\llm} ({\method{o1-preview}}, the most advanced OpenAI
model at the time of experimentation) to obtain the version shown in
Figure~\ref{quote:prompt_task_1}.
The refined prompt incorporates placeholders for demographic
information to be inserted (gender and age group), allowing
exploration of how {\llms} judge sexism in tweets when instructed to
do so from different perspectives.

%------------------------------------------
\input{prompts/LLM_generated_base}
%------------------------------------------

%------------------------------------------
\myparagraph{Bias Analysis}
%------------------------------------------

To study whether {\llm} predictions are subject to bias, a series of
evaluations were conducted.
Model predictions were generated first using the baseline prompt
without any demographic cues (that is, without the bold text shown in
Figure~\ref{quote:prompt_task_1}).
The two demographic factors -- gender and age -- were then
incorporated into the prompt, to explore their impact on sexism
classification.
Agreement between model outputs and human annotations was quantified
using Krippendorff's $\alpha$.
To establish the robustness of the analysis, we conducted confidence
intervals using bootstrap resampling ($10{,}000$ iterations).
All of the measured Krippendorff's $\alpha$ coefficients had
confidence intervals smaller than $0.001$.

%------------------------------------------
\input{Tables/tbl_t1_kripp_alpha_gender}

\input{Tables/tbl_t1_kripp_alpha_age}
%------------------------------------------

%% file: Tables/tbl_exist2023_distribution.tex
\begin{table}[tb]    \centering
    \mycaption{{\method{EXIST}} 2023 label distribution over $7{,}958$ tweets.}
    \label{tab:exist2023_data_distribution}
    \begin{tabular}{lccccc}
        \toprule
        & \multicolumn{2}{c}{\textbf{Gender}} & \multicolumn{3}{c}{\textbf{Age}} \\
        \cmidrule(lr){2-3} \cmidrule(lr){4-6}
        & \textbf{Female} & \textbf{Male} & \textbf{18--22} & \textbf{23--45} & \textbf{46+} \\
        \midrule
        \text{Sexist} & 10,961 & 10,790 & 6,933 & 7,422 & 7,396 \\
        \text{Non-Sexist} & 12,913 & 13,084 & 8,983 & 8,494 & 8,520 \\
        \midrule
        \text{Total} & 23,874 & 23,874 & 15,916 & 15,916 & 15,916 \\
        \bottomrule
    \end{tabular}
\end{table}

%% file: prompts/LLM_generated_base.tex
\begin{figure}[t]
%\begin{quotation}
\begin{minipage}{8.5cm}
\begin{lstlisting}[frame=single,numbers=none,basicstyle=\footnotesize\ttfamily]
You are an expert linguist specializing in detecting 
sexism in social media texts. (*@\textbf{Your demographic information}@*)
(*@\textbf{is as follows:}@*)
(*@\textbf{-Sex: [insert sex: male/female]}@*)
(*@\textbf{-Age Group: [insert age group: 18-22/23-45/46+]}@*)

Task: Analyze the given tweet and determine whether it 
contains sexist expressions or behaviors. This includes 
tweets that:
-Are sexist themselves (contain sexist language or 
promote sexist ideas).
-Describe a sexist situation (narrate or depict scenarios 
where sexism occurs).
-Criticize sexist behavior (call out or oppose sexism).

Instructions:
1. Read the Tweet Carefully: Examine the content of the 
tweet thoroughly.
2. Make a Determination: Decide if the tweet fits any of 
the above categories related to sexism.
3. Classification:
-Assign "YES" if the tweet contains any form of sexist 
content as defined.
-Assign "NO" if the tweet does not contain sexist content.
4. Output Format: Provide the assigned Category in plain
text.
5. Constraint: You must not retrieve any text apart from 
the two possible categories, YES and NO.

TWEET: [insert tweet]
\end{lstlisting}
\end{minipage}
\captionsetup{skip=5pt} %reduces space between label and table
\mycaption{Prompt structure.
Bold text indicates the parts that were added/varied in the various
experiments, as extensions beyond the ``baseline'' prompt.
%\falk{Is this verbatim what was used? Did we really end up with the word ``retrieve'' in item 5? :-)} \jt{I think we did...}
}
\label{quote:prompt_task_1}
%\end{quotation}
\end{figure}

%% file: Tables/tbl_t1_kripp_alpha_gender.tex
% \begin{table}[h]
%     \centering
%     \mycaption{Correlation Scores Between Different Models and Gender Groups}
%     \label{tab:correlation}
%     \begin{tabular}{lcc}
%         \toprule
%         \textbf{} & \textbf{F} & \textbf{M} \\
%         \midrule
%         F & \textbf{1.000} & 0.477 \\
%         M & 0.477 & \textbf{1.000} \\
%         \midrule
%         gpt-3.5 & \textbf{0.415} & 0.371 \\
%         gpt-3.5 + female & \textbf{0.398} & 0.358 \\
%         gpt-3.5 + male & \textbf{0.404} & 0.360 \\
%         \midrule
%         gpt-4 & \textbf{0.365} & 0.325 \\
%         gpt-4 + female & \textbf{0.401} & 0.360 \\
%         gpt-4 + male & \textbf{0.372} & 0.336 \\
%         \midrule
%         gpt-4o & \textbf{0.228} & 0.191 \\
%         gpt-4o + female & \textbf{0.234} & 0.198 \\
%         gpt-4o + male & \textbf{0.213} & 0.172 \\
%         \bottomrule
%     \end{tabular}
% \end{table}
\begin{table}[tb]
    \centering
    \mycaption{Krippendorff's $\alpha$ scores comparing human annotators by gender with LLMs. Bold text indicates the model's highest agreement across all gender groups. Subscripts F and M denote gender-based personas through persona prompting in {\llms}.
    % For instance, \text{GPT-3.5} showed the highest agreement with the Female group ($\alpha=0.415$) compared to the Male group.
    }
    \label{tab:krippendorff_alpha}
    \begin{tabular}{lcc}
        \toprule
        \textbf{Model} & \textbf{F (Female)} & \textbf{M (Male)} \\
        \midrule
        Human Annotators (F) & \textbf{1.000} & 0.477 \\
        Human Annotators (M) & 0.477 & \textbf{1.000} \\
        \midrule
        \gptthree & \textbf{0.415} & 0.371 \\
        \gptthreefemale & \textbf{0.398} & 0.358 \\
        \gptthreemale & \textbf{0.404} & 0.360 \\
        \midrule
        \gptfour & \textbf{0.365} & 0.325 \\
        \gptfourfemale & \textbf{0.401} & 0.360 \\
        \gptfourmale & \textbf{0.372} & 0.336 \\
        \midrule
        \gptl & \textbf{0.228} & 0.191 \\
        \gptlfemale & \textbf{0.234} & 0.198 \\
        \gptlmale & \textbf{0.213} & 0.172 \\
        \midrule
        \mistral & \textbf{0.353} & 0.310 \\
        \mistralfemale & \textbf{0.363} & 0.326 \\
        \mistralmale & \textbf{0.330} & 0.293 \\
        \midrule
        \qwen & \textbf{0.378} & 0.345 \\
        \qwenfemale & \textbf{0.372} & 0.337 \\
        \qwenmale & \textbf{0.382} & 0.347 \\
        \bottomrule
    \end{tabular}
\end{table}

%% file: Tables/tbl_t1_kripp_alpha_age.tex
\begin{table}[tb]
    \centering
    \mycaption{Krippendorff's $\alpha$ scores comparing human annotators by age group with {\llms}. Bold text indicates the model's highest agreement across all age groups. Subscripts denote age-based personas introduced through prompting in {\llms}.
    % For instance, \text{GPT-3.5} exhibited the highest agreement with the 46+ age group ($\alpha=0.413$) compared to the 18–22 and 23–45 age groups.
    }
    \label{tab:krippendorff_alpha_age}
    \begin{tabular}{lccc}
        \toprule
        \textbf{Model} & \textbf{18--22} & \textbf{23--45} & \textbf{46+} \\
        \midrule
        Human Annotators (18--22) & \textbf{1.000} & 0.445 & 0.436 \\
        Human Annotators (23--45) & 0.445 & \textbf{1.000} & 0.463 \\
        Human Annotators (46+) & 0.436 & 0.463 & \textbf{1.000} \\
        \midrule
        \gptthree & 0.382 & 0.408 & \textbf{0.413} \\
        \gptthreeageone & 0.372 & 0.399 & \textbf{0.409} \\
        \gptthreeagetwo & 0.365 & 0.398 & \textbf{0.402} \\
        \gptthreeagethree & 0.383 & 0.407 & \textbf{0.419} \\
        \midrule
        \gptfour & \textbf{0.421} & \textbf{0.421} & 0.404 \\
        \gptfourageone & 0.455 & \textbf{0.462} & 0.452 \\
        \gptfouragetwo & 0.446 & \textbf{0.484} & 0.430 \\
        \gptfouragethree & 0.463 & \textbf{0.474 }& 0.457 \\
        \midrule
        \gptl & \textbf{0.316} & 0.290 & 0.278 \\
        \gptlageone & \textbf{0.286} & 0.261 & 0.247 \\
        \gptlagetwo & \textbf{0.302} & 0.272 & 0.265 \\
        \gptlagethree & \textbf{0.302} & 0.271 & 0.262 \\
        \midrule
        \mistral & 0.368 & 0.384 & \textbf{0.392} \\
        \mistralageone & 0.372 & 0.389 & \textbf{0.392} \\
        \mistralagetwo & 0.378 & 0.392 & \textbf{0.398} \\
        \mistralagethree & 0.360 & 0.377 & \textbf{0.383} \\
        \midrule
        \qwen & 0.406 & \textbf{0.418} & 0.404 \\
        \qwenageone & 0.421 & \textbf{0.432} & 0.424 \\
        \qwenagetwo & 0.423 & \textbf{0.437} & 0.427 \\
        \qwenagethree & 0.412 & \textbf{0.419} & 0.411 \\        
        \bottomrule
    \end{tabular}
\end{table}

%% file: sec4-results.tex
%------------------------------------------
\section{Results}
\label{sec-results}
%------------------------------------------

\smallskip

\myparagraph{Do {\llms} exhibit bias toward certain demographics when
classifying text as sexist or non-sexist?}

We evaluate Krippendorff's \text{$\alpha$} between each {\llm} and
human annotators across two demographics -- gender and age.

First, we compare Krippendorff's $\alpha$ of {\llms} against female
and male annotators.
Table~\ref{tab:krippendorff_alpha} shows that {\llms} consistently
show higher agreement with female annotators than with male
annotators when classifying text as sexist, highlighting an inherent
gender-based bias, with {\gptthree} demonstrating the highest
agreement with female annotators, and {\gptl} the lowest.
Additionally, as we progress from {\gptthree} to {\gptfour} and
subsequently to {\gptl}, we observe a decrease in agreement with both
male and female annotators, indicating a decline in alignment as the
models have evolved.

On the other hand, age-related patterns do not follow a
consistent trend of bias across {\llms}.
Table~\ref{tab:krippendorff_alpha_age} shows that {\gptthree} and
{\mistral} align more closely with annotators aged $46$+, while
{\gptfour} and {\qwen} show higher agreement with the $23$--$45$ age
group.
Meanwhile, {\gptl} aligns most with annotators aged $18$--$22$.

Overall, our results demonstrate that {\llms} exhibit bias towards
certain demographics when classifying text as sexist, with higher
agreement with female annotators than with male annotators.
Note that throughout these results the observed differences were
greater than the computed confidence intervals of the measured values.

\myparagraph{Can adopting a demographic-based persona mitigate bias
in {\llms} when classifying text as sexist or non-sexist?}

Demographic factors -- specifically gender and age -- were
incorporated into the {\llm} prompt to mitigate biases by adopting a
demographic-based persona, indicated by the bold text sections in
Figure~\ref{quote:prompt_task_1}.

Table~\ref{tab:krippendorff_alpha} shows that only {\gptfourfemale},
{\gptlfemale}, and {\mistralfemale} exhibited increased agreement
with female annotators compared to their base models, where the
subscripts indicate the instruction added to the {\llm} prompts.
However, {\gptthreefemale} demonstrated a decrease in agreement with
female annotators.
Similarly, only {\gptfourmale} and {\qwenmale} demonstrated improved
agreement with male annotators, while remaining {\llms} showed a
decline.
These findings indicate that incorporating gender-based personas in
prompts should not be assumed to mitigate gender bias in {\llms} for
this task.

For age-based personas, Table~\ref{tab:krippendorff_alpha_age} shows
that {\gptfour}, {\mistral}, and {\qwen} consistently exhibited
higher agreement with the prompt-included age grouping than the base
models, whereas {\gptthree} demonstrated increased agreement only for
the $46$+ age group.
Meanwhile {\gptl} did not indicate any improvement in agreement along
any age groups.

Overall, the findings indicate that instructing {\llms} to adopt
demographic-based personas has inconsistent and unpredictable
effects.
While persona prompting improves alignment with certain demographic
groups in some models, it decreases alignment in others, meaning that
it cannot be relied upon, and that it may not be presumed to provide
a reliable way of mitigating bias.

%% file: sec5-conclusion.tex
%------------------------------------------
\section{Discussion and Conclusion}
\label{sec:discussion}
%------------------------------------------

In this study, we investigated demographic biases in {\llms} when
classifying text as sexist or non-sexist.
By analyzing Krippendorff’s $\alpha$ agreement between {\llm}-based
annotations and human annotations across gender and age breakdowns,
we found that {\llms} consistently align more closely with female
annotators than male annotators, indicating a gender-based bias.
However, age-based breakdowns exhibited no clear pattern of bias,
with different models aligning with different age groups.

Adopting a structure that might be thought of as addressing these
biases, we also explored the effectiveness of persona prompting,
incorporating gender and age information into the {\llm} prompts.
Our results indicate that this approach yields inconsistent and
unpredictable outcomes.
Some models improved alignment with specific demographic groups, but
others showed decreased agreement.
We thus cannot (yet) regard demographic-based prompting to be an
effective way of mitigating {\llm} bias, an outcome that we expect
will be of interest (and concern) to other researchers.

%------------------------------------------
\myparagraph{Future Work}
%------------------------------------------

We plan to extend our study beyond binary classification to explore
whether incorporating more granular categories, such as those defined
by {\citet{plaza2023exist}}, can provide insights into the
inconsistencies observed in model agreement.
Moreover, humans possess multi-faceted personas that encompass the
intersectionality of various demographics {\cite{liu2024evaluating}}.
While the current study addresses one-dimensional personas, we aim to
explore demographic intersectionality in future work by examining how
multiple factors interact and whether this results in more consistent
alignment across {\llms}.
Beyond age and binary gender, other characteristics may also shape
annotators’ perspectives on sexism, and we plan to assess their
influence on model behavior.
Lastly, we plan to extend our study to a wider range of {\llms}, to
identify what common factors may exist, and to assess the
generalizability of our findings.